\documentclass[aps,pra,showpacs,floatfix,onecolumn]{revtex4}
\usepackage[letterpaper,margin=2cm]{geometry}
\usepackage[brazil,english]{babel}
\usepackage[T1]{fontenc}
\usepackage{amsmath,amsfonts,amssymb,float,graphicx,epsfig,color,times,bbm}

\usepackage{hyperref}
\begin{document}
\title{Creation and localization of entanglement in a simple configuration of coupled harmonic oscillators}
\author{J. F. Leandro}
\affiliation{Departamento de Física, Universidade Estadual de Ponta Grossa - Campus Uvaranas, Ponta Grossa, Paraná 84030-900, Brazil}
\author{F. L. Semi\~ao}
\affiliation{Departamento de Física, Universidade Estadual de Ponta Grossa - Campus Uvaranas, Ponta Grossa, Paraná 84030-900, Brazil}
\begin{abstract}
 We investigate a simple arrangement of coupled harmonic oscillators which brings out some
interesting effects concerning creation of entanglement. It is well known that if each member in a linear chain of coupled harmonic oscillators is
prepared in a ``classical state'', such as a pure coherent state or a mixed thermal state, no entanglement is created
in the rotating wave approximation. On the other hand, if one of the oscillators is prepared in a nonclassical state (pure squeezed state, for instance), entanglement may be created between members of the chain. In the setup considered here, we found that a great family of nonclassical (squeezed) states can localize entanglement in such a way that distant oscillators never become entangled. We present a detailed study of this particular localization phenomenon. Our results may find application in future solid state implementations of quantum computers, and we suggest an electromechanical system consisting of an array of
coupled micromechanical oscillators as a possible implementation.  
\end{abstract}
\maketitle
Nonclassicality of quantum radiation fields is an important topic in quantum optics \cite{loudon,lknight,mandel,hillery,lee,dodonov,kim,richter}. A single-mode radiation field manifests nonclassical effects when it is in a state whose \textit{P} function (also known as coherent state representation or Glauber-Sudarshan Distribution) is quite singular and not positive definite \cite{lee}. General measures of nonclassicality have been proposed in the quantum optics literature \cite{hillery,lee,dodonov,kim,anatole}. An important class of field states comprises those states whose Wigner function is Gaussian \cite{marian1}. These Gaussian states attracted much attention since they can be created and manipulated with reasonably simple optical elements and because measures of nonclassicality are also known for them \cite{marian2,marian3,janos}.

Equally important in quantum optics, and more recently in quantum-information science, is the concept of entanglement, a typical quantum mechanical phenomenon which has been investigated since the beginning of the quantum theory \cite{epr,bell}. Nonclassical properties, particularly squeezing and creation of entanglement, seem to be linked concepts at least when initially product squeezed states are manipulated with passive Gaussian operations \cite{wolf}. In fact, among the different measures of nonclassicality of a field state, there is one which uses the degree of entanglement generated from separable Gaussian states when assisted by beam splitters, phase shifters, auxiliary classical fields, and ideal photodetectors \cite{janos}. Actually, beam splitters have become a useful tool for entangling light fields, and its capacity as so has been extensively studied \cite{bsent,li,tahira}. Much is known about nonclassicality of input states and entanglement in the output states of a beam splitter, but more general setups are still considerably unexplored. Since a system of two coupled harmonic oscillators in the rotate wave approximation (RWA) is, under certain conditions, isomorphic to two field modes coupled by a beam splitter, such general setups seem to be more easily visualized when considering arrangements of harmonic oscillators corresponding to arbitrary weighted graphs. This paper is intended to investigate the effects of squeezing on the creation of entanglement in a simple arrangement of coupled harmonic systems. We will show that for a great range of the relative direction of squeezing of two reference oscillators, entanglement becomes localized, and no long distance entanglement is created whatsoever.

\begin{figure}[ht]
 \centering\includegraphics[width=0.55\columnwidth]{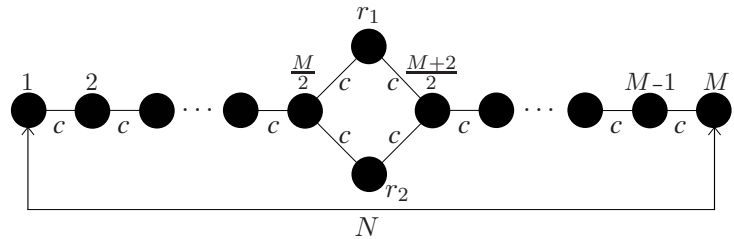}
 \caption{A diagram of the interferometric setup used in this paper. Identical oscillators (unit mass and frequency) are initially prepared in the vacuum state except for two reference oscillators (labeled with $r_1$ and $r_2$) which are initially prepared in pure squeezed states.  $N$ is the logarithmic negativity (base $e$) calculated between the oscillators at the extremes of the chains and $c$ are coupling constants.}
 \label{fig1}
\end{figure}

The dynamics of entanglement in systems consisting of coupled harmonic oscillators has been studied in great detail in \cite{martin}. In addition to a careful study of the propagation of entanglement in linear chains, more general geometrical arrangements of the harmonic oscillators have also been considered in \cite{martin}. Among these more general arrangements, we will be considering the setup depicted in Fig.\ref{fig1}.
This setup may be seen as a small ring coupled to two chains. In our case, this ring consists of four oscillators, where two of them are used to allow coupling with the chains (hubs) and the other two are reference oscillators where product pure squeezed states may be prepared. With exception of these two reference oscillators, the rest is initially prepared in a product of individual vacuum states. The RWA Hamiltonian for the system depicted in Fig.\ref{fig1} is given by ($\hbar=m=\omega=1$)

\begin{eqnarray}\label{H}
H&=&\frac{1}{2}(p_{r_1}^2+q_{r_1}^2)+\frac{1}{2}(p_{r_2}^2+q_{r_2}^2)+\frac{1}{2}\sum_{i=1}^M\left(p_{i}^2+q_{i}^2\right)+\frac{c}{4}(q_{r_1}-q_{\frac{M}{2}})^2+\frac{c}{4}(p_{r_1}-p_{\frac{M}{2}})^2\nonumber\\ &&+ \frac{c}{4}(q_{r_1}-q_{\frac{M+2}{2}})^2+ \frac{c}{4}(p_{r_1}-p_{\frac{M+2}{2}})^2+ \frac{c}{4}(q_{r_2}-q_{\frac{M}{2}})^2+ \frac{c}{4}(p_{r_2}-p_{\frac{M}{2}})^2\nonumber\\ &&+ \frac{c}{4}(q_{r_2}-q_{\frac{M+2}{2}})^2+ \frac{c}{4}(p_{r_2}-p_{\frac{M+2}{2}})^2+
\frac{c}{4}\sum_{{i=1}\atop{i\neq \frac{M}{2}}}^{M-1}\left[(q_{i+1}-q_{i})^2+(p_{i+1}-p_{i})^2\right].
\end{eqnarray}
Hamiltonian (\ref{H}) can be put into the suitable form
\begin{eqnarray}
\hat{H}=\frac{1}{2}\sum_{ij}(p_iT_{ij}p_j+q_iV_{ij}q_j)=\frac{1}{2}R\left[ \begin{array}{cc} V & 0\\ 0 & T\end{array}\right]R^T,\label{Ho}
\end{eqnarray}
where $R=(q_{r_1},q_{r_2},q_1,\dots,q_M,p_{r_1},p_{r_2},p_1,\ldots,p_M)^T$ is a vector grouping position and momentum operators and $V$ and $T$ are symmetric matrices whose elements can be found from Eq.(\ref{H}). Note that $T=V$ because we are working in the RWA.

Since Hamiltonian (\ref{Ho}) is quadratic in the position and momentum operators, initial Gaussian states will remain Gaussian at all times, and this facilitates the study of entanglement because a computable entanglement measure is known for such states \cite{logneg}. Entanglement properties of Gaussian states are contained in the covariance matrix whose elements are given by
\begin{equation}
\gamma_{ij}=2\,{\rm{Re}}\,{\rm{tr}}[\rho(R_i-\langle
R_i\rangle)(R_j-\langle R_j\rangle)],\label{cov}
\end{equation}
where the expectation values are evaluated with the system density operator $\rho$. For a bipartite system composed of subsystems $A$ and $B$, an entanglement measure called logarithmic negativity can be written as a function of the covariance matrix as \cite{logneg}
\begin{equation}
N=-\sum_j\ln[\min(1,|\gamma_j^{T_B}|)],\label{logsym}
\end{equation}
where $\gamma_j^{T_B}$ are the symplectic eigenvalues of the covariance matrix $\gamma^{T_B}$, evaluated from the system density operator after partial transposition of system $B$. Actually, there is no need to work with the system density operator since the effect of partial transposition of $\rho$ on the covariance matrix is easily obtained from the transformation $\gamma^{T_B}=\mathcal{P}\gamma \mathcal{P}$, where for a bipartite system consisting of two oscillators \cite{martin}
\begin{eqnarray}
\mathcal{P}=\left[ \begin{array}{cccc} 1 & 0 & 0 & 0\\ 0 & 1 & 0 & 0 \\ 0 & 0 & 1 & 0 \\ 0 & 0 & 0 & -1\end{array}\right].
\end{eqnarray}
The time evolution of the covariance matrix for quadratic Hamiltonian (\ref{Ho}) is given by the simple expression \cite{martin}
\begin{eqnarray}
\left[ \begin{array}{cc} \gamma_{XX}(t) & \gamma_{XP}(t)\\ \gamma_{PX}(t) & \gamma_{PP}(t)\end{array}\right]=E(t)\left[ \begin{array}{cc} \gamma_{XX}(0) & \gamma_{XP}(0)\\ \gamma_{PX}(0) & \gamma_{PP}(0)\end{array}\right]E^\dag(t)\label{ev}
\end{eqnarray}
where
\begin{eqnarray}
E(t)=\exp\left[\left(\begin{array}{cc} 0 & T\\ -V & 0\end{array}\right)t\right].
\end{eqnarray}
We now have all the tools needed to study the dynamics of entanglement in this system.

It is a well known fact that initial ``classical states'', such as coherent or thermal states, are not capable of generating entanglement in a linear chain of RWA-coupled harmonic oscillators. Many papers have then studied the problem of the entangling capacity of RWA coupling and its dependence on the initial state of the oscillators \cite{bsent,li,tahira}. Although nonclassical states are needed for generating entanglement using this coupling, not all nonclassical states lead to entanglement. Two RWA-coupled harmonic oscillators (with lowering operators $a$ and $b$) which are initially prepared in pure squeezed states $|r_ae^{i\phi_a}\rangle$ and $|r_be^{i\phi_b}\rangle$, respectively, will never become entangled with the interaction $H=g(a^\dag b+b^\dag a)$ when $r_a=r_b$ and relative angle $\phi_b-\phi_a=\pi$. In this case, there is just one relative direction of squeezing in the domain $[0,2\pi)$ which does not generate entanglement, namely $\phi_b-\phi_a=\pi$.

We will now see that the setup depicted in Fig.\ref{fig1} presents a very rich structure concerning the dependence of the dynamics of entanglement upon relative direction of squeezing of the two reference oscillators. We will show that there is a great range of the relative direction of squeezing of the two reference oscillators $r_1$ and $r_2$ for which no entanglement between the oscillators $1$ and $M$ is created whatsoever. As mentioned before, with exception of the two reference oscillators, all other oscillators are initially prepared in individual vacuum states. Preparation of squeezed states of linear coupled harmonic oscillators outside RWA has been studied in \cite{tag}, where a time dependent localization of entanglement also appears. We will show that for the interferometric setup treated here, perfect localization of entanglement occurs and it is time independent for a great family of squeezed states.

\begin{figure}[ht]
 \centering\includegraphics[width=0.55\columnwidth]{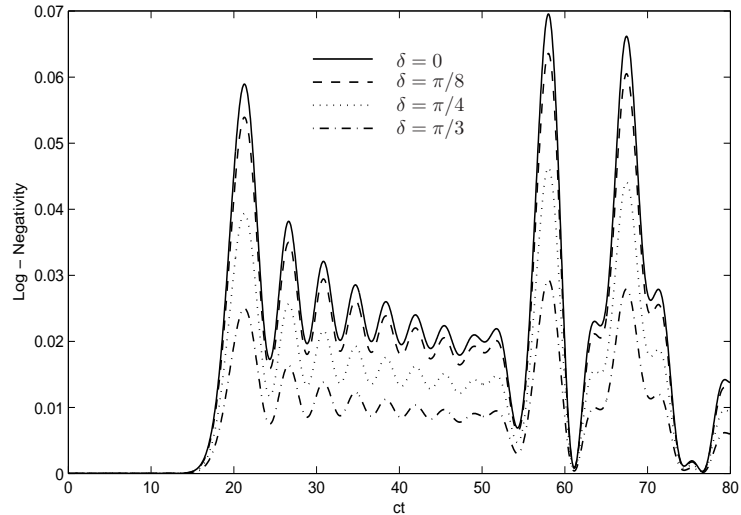}
 \caption{Time evolution of the logarithmic negativity for different relative angles of squeezing. We have used $M=38$ and $r=1$.}
 \label{fig2}
\end{figure}

\begin{figure}[ht]
 \centering\includegraphics[width=0.45\columnwidth]{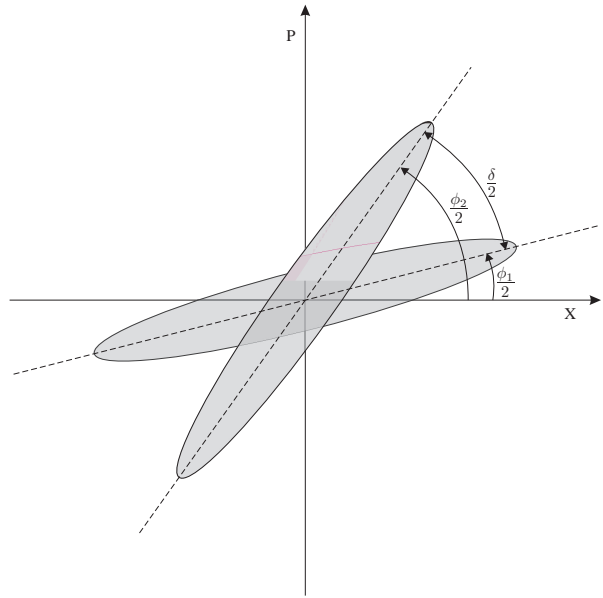}
 \caption{Phase-space representation of the two squeezed states considered in this paper. The reference oscillators are prepared in squeezed states $|r e^{i\phi_{1}}\rangle$ and $|r e^{i\phi_{2}}\rangle$ with same squeezing but with a general relative angle of squeezing $\delta=\phi_2-\phi_1$.}
 \label{fig3}
\end{figure}

\begin{figure}[ht]
 \centering\includegraphics[width=0.55\columnwidth]{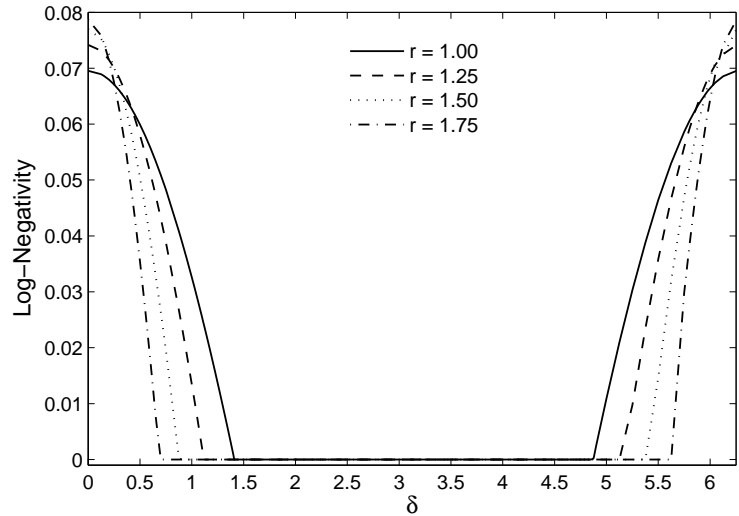}
 \caption{Logarithmic negativity at $ct=58$ as a function of the relative angle of squeezing $\delta$ in the case $M=38$.}
 \label{fig4}
\end{figure}

In Fig.\ref{fig2}, we plot the time evolution of the entanglement between the oscillators at each extreme of the chain considering $M=38$. The reference oscillator $r_1$ is initially prepared in the pure squeezed state $|r_1 e^{i\phi_{1}}\rangle$ and the reference oscillator $r_2$ is prepared in the pure squeezed state $|r_2 e^{i\phi_{2}}\rangle$, with $r_1=r_2=r$. This initial preparation was used for all plots shown in this paper. It is interesting to notice that for any fixed time, the entanglement decreases as the relative angle of squeezing $\delta\equiv\phi_2-\phi_1$ increases. This relative angle of squeezing is depicted in Fig.\ref{fig3}.\vspace{0.15cm}
As $\delta$ increases, it comes to a point that no entanglement is created between oscillators $1$ and $M$ at any instant of time. This time independence was verified numerically for many different values of $M$. We provide an analytical exact result for $M=2$ later on this paper. In contrast to the case of just two RWA-coupled oscillators, $\delta$ is not unique in the interval $[0,2\pi)$. Actually, there is now a continuum of values of $\delta$ which forbids the creation of entanglement. As one keeps increasing $\delta$, it comes to a point where entanglement starts arising again. This range of $\delta$ which forbids creation of entanglement between $1$ and $M$ depends on the squeezing parameter $r$ for the initial preparation we have been considering so far. In Fig.\ref{fig4}, we present the variation in $\delta$ as a function of the squeezing parameter $r$ for a fixed time $ct=58$. This time is approximately the first local maximum of the entanglement for $ct<80$ as one can see in Fig.\ref{fig2}. It is clearly shown in Fig.\ref{fig4} that the increasing in the squeezing parameter $r$ leads to the increasing in the range of $\delta$ for which no entanglement can be created. This clearly indicates that the statement that squeezing favors entanglement creation has to be seen in context. Here, we found an example that the more squeezed the reference oscillators, the more resistant to entanglement creation the setup becomes in the particular sense that the range of values of $\delta$ allowing entanglement creation between $1$ and $M$ decreases. As $r$ increases, the system becomes then more selective concerning entanglement generation between the extremes of the chain.

\begin{figure}[ht]
 \centering\includegraphics[width=0.55\columnwidth]{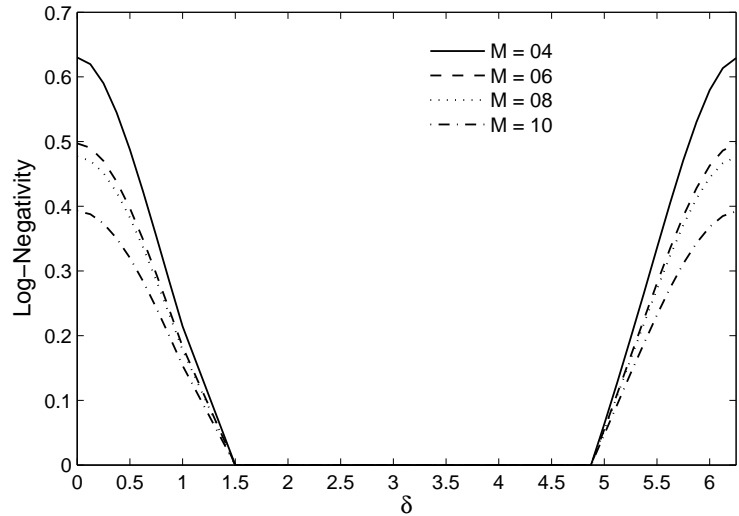}
 \caption{First local maximum of the logarithmic negativity in the domain $ct<80$ as a function of the relative angle of squeezing $\delta$ for different numbers $M$ of oscillators in the chains. We considered $r=1$.}
 \label{fig5}
\end{figure}

\begin{figure}[ht]
 \centering\includegraphics[width=0.55\columnwidth]{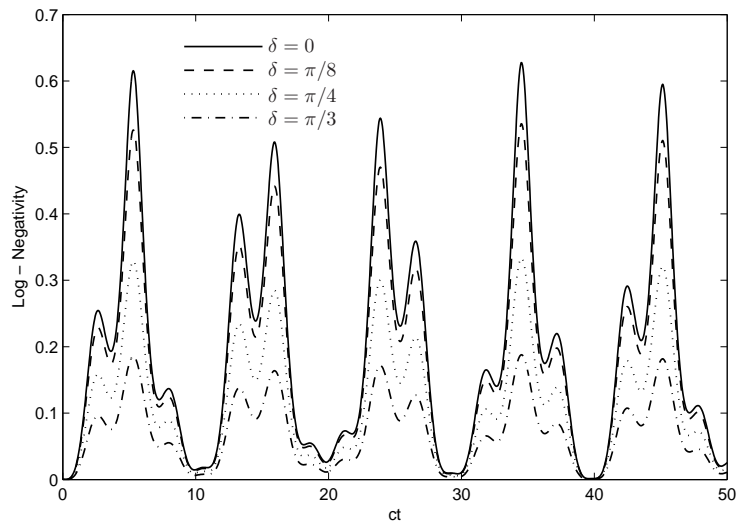}
 \caption{Time evolution of the logarithmic negativity for $M=4$ and different relative angles of squeezing $\delta$. We considered $r=1$.}
 \label{fig6}
\end{figure}

The reason why entanglement does not propagate in the situation just described seems to reside in the fact that the entanglement between the oscillators $M/2$ and $(M+2)/2$ is zero for the range of $\delta$ which forbids creation of entanglement between $1$ and $M$. Entanglement is created just between $M/2$, $r_1$ and $r_2$, and also between $(M+2
)/2$, $r_1$ and $r_2$, but no bipartite entanglement between $M/2$ and $(M+2)/2$ exists for such a range of $\delta$. This localized entanglement persists for all times. From the point of view of propagation of entanglement, there is no entanglement being propagated to oscillators $1$ and $M$ because the hubs $M/2$ and $(M+2)/2$ are not entangled for all times. It is clearly displayed in Fig.\ref{fig2} that the oscillators $1$ and $M$ take time to get entangled. This is also in accordance with the view of entanglement being created in the center of the setup (in the ring), and then being propagated. Propagation of entanglement is also carefully studied in \cite{martin,martin2}
It is worthwhile to notice that the range of values of $\delta$ seems to be independent of the number of oscillators in the chains. We numerically tested this conjecture for $M$ ranging from $2$ to $98$. In Fig.\ref{fig5}, we show this independence for some values of $M$. In this plot, we considered the maximum entanglement for times $ct<80$, but another choices would lead to similar results since the localization is time-independent. As one could expect, the entanglement decreases with the distance (number of oscillators in the chains).
Since our results seemed to be independent of $M$, we exactly solved the problem for the ring only (this situation corresponds to $M=2$). In fact, the essence of the entanglement localization lies in the properties of the ring. The only symplectic eigenvalue that can be smaller than $1$ in the case $M=2$ reads
\begin{eqnarray}
\gamma^{T_B}=\{|\cos^2(c t)+\sin^2(c t)[\cosh(2r)-\sinh(2r)|\cos(\delta/2)|]|\}^{1/2}.
\end{eqnarray}
One can easily see that in order for $\gamma^{T_B}$ to be smaller than $1$ (creation of entanglement between oscillators $1$ and $2$), the following inequality must be satisfied for
\begin{eqnarray}
\left|\cos\left(\frac{\delta}{2}\right)\right|<\frac{\cosh(2r)-1}{\sinh(2r)}.\label{in}
\end{eqnarray}
It is worthwhile to notice that this inequality does not depend on the coupling constant $c$. For $r=1$, one can see that inequality (\ref{in}) is satisfied for $-1.41005\leq \delta\leq 1.41005$. Then, if $1.41005\leq\delta\leq 4.87313$, the oscillators $1$ and $M$ $(M=2)$ will not get entangled. This is in accordance with numerics for $M=38$ shown in Fig.\ref{fig4}. Applying now Eq.(\ref{in}) for $r=1.75$, entanglement is created between those oscillators if $-0.68822\leq \delta\leq 0.68822$. Consequently, if $0.68822\leq\delta \leq 5.59496$ oscillators $1$ and $38$ will not get entangled. Again, the case with more oscillators shown in Fig.\ref{fig4} agrees completely with the simple setup $M=2$. Although the dependence of $\delta$ on $r$ and $c$ is independent of the number of oscillators, the amount of entanglement and its dynamics changes completely with $M$. As an example, we show in Fig.\ref{fig6} the time evolution of the entanglement between the oscillators $1$ and $M$ for $M=4$. When compared with the plot shown in Fig.\ref{fig2}, we can see that the frequencies involved in the evolution and the amount of entanglement available for the two oscillators at the extremes of the chain change with $M$. However, the localization as a function of $\delta$ will be the same.

\begin{figure}[ht]
 \centering\includegraphics[width=0.55\columnwidth]{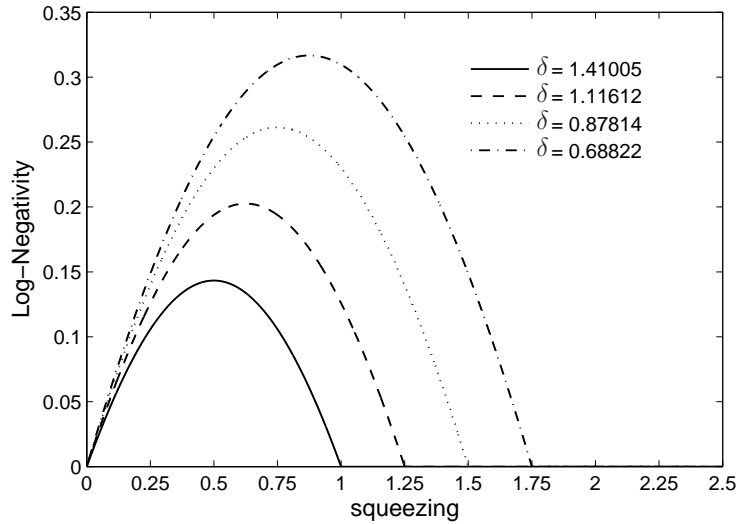}
 \caption{Logarithmic negativity at $ct=1$ as a function of the magnitude of squeezing and for different relative angles $\delta$. We considered $M=2$.}
 \label{fig7}
\end{figure}

\begin{figure}[ht]
 \centering\includegraphics[width=0.55\columnwidth]{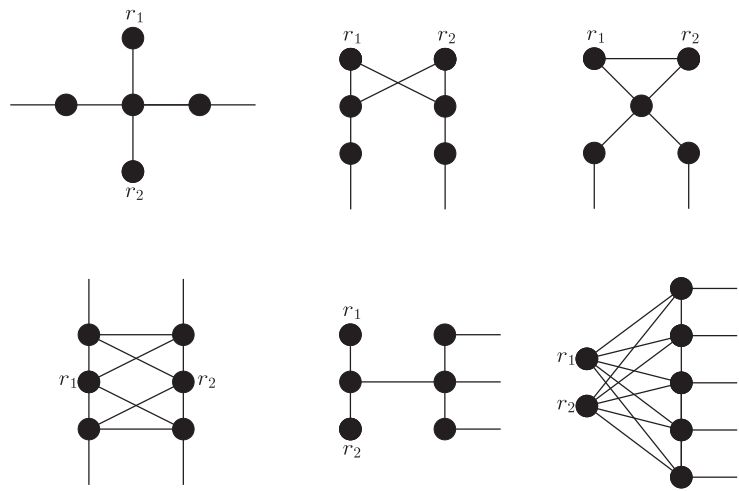}
 \caption{Examples of different configurations (structures) which localize entanglement. Chains of arbitrary length can be coupled to such structures through the open lines (hubs) in order to study entanglement creation and propagation. Depending on the relative angle of squeezing of the two reference oscillators $r_1$ and $r_2$, no entanglement will be generated between the members of the chains coupled to these structures.}
 \label{fig8}
\end{figure}

It may be interesting to look at this relation between $\delta$ and the entanglement from a different point of view, i.e., by fixing $\delta$ and changing $r$. In Fig.\ref{fig7}, we show how the entanglement between $1$ and $M$, for the simple case $M=2$, depends on the squeezing parameter for different values of $\delta$. It is clear that once $\delta$ is fixed, there is an upper bound for $r$. Above that upper bound, no entanglement is generated regardless the fact that the more squeezed the initial state is, the more nonclassical it becomes. The results shown in Figs.\ref{fig4}, \ref{fig5}, and \ref{fig7} are consequences from inequality (\ref{in}).
It is now important to make a brief comment about the choice of the system depicted in Fig.\ref{fig1}. Our primary aim in this paper is to show the strong dependence of the perfect localization of entanglement on the phase difference of two squeezed reference oscillators.  Chains of arbitrary length are used in order to emphasize the localization of entanglement in the central structure. We realized that this phenomenon occur when at least one further oscillator is coupled to both reference oscillators. We have chosen the configuration depicted in Fig.\ref{fig1} since it is the simplest situation one can have where at least two hubs are available for further connections. More involved choices are displayed in Fig.\ref{fig8}.
Finally, we would like to connect our study with applications in a solid state setup. Nowadays, microelectromechanical systems (MEMSs) can be manufactured and controlled with impressive precision. If cooled down to sufficiently low temperatures, these systems would operate as harmonic oscillators in the quantum regime. The coupling constants between neighbor MEMS are controlled by means of voltages biases. Experimental realization of a large array of doubly clamped beams made of gold fabricated on top of a silicon nitrite membrane is reported in \cite{exp}. Previous theoretical proposals studying entanglement of coupled harmonic oscillators have also suggested this electromechanical setup as a potential candidate for physical implementation \cite{nano,multib}. Decoherence lowers the \textit{Q} factors for the MEMS, and values of $Q=10^3-10^4$ are realistic nowadays \cite{exp,multib}. According to \cite{nano}, most of the dissipation and decoherence is expected
to come from the coupling of each gold beam with the degrees of freedom of the silicon substrate to which the resonator is connected to (phononic modes). This leads to a simple Gaussian decoherence model for noise in these coupled MEMSs \cite{nano}. It is shown in \cite{nano} that entanglement creation and propagation are surprisingly robust against noise caused by the coupling with the substrate. Indeed, eight MEMS oscillators with natural frequencies of $5$ GHz and cooled down to temperatures about 10 mK would dynamically lead to logarithmic negativities about $0.2$ for $Q=10^3$ and $c=0.3$ in the scheme presented in \cite{nano}. This is a strong motive for quoting MEMS reported in \cite{exp} as a possible future implementation of the ideas presented here.

To summarize, we have studied the dynamics of entanglement in a simple arrangement of coupled harmonic oscillators with special emphasis on the relation between squeezing and creation of entanglement. We have found that for a great family of squeezed states for two reference oscillators there is a strong localization of entanglement. We performed a careful analysis of the parameters involved, and concluded that both the magnitude of the squeezing parameters and their phases are important in the creation of entanglement in our setup. We have shown that the main features of entanglement localization for an arbitrary number of oscillators in our setup depend only on the central oscillators (ring) and could be reproduced in the simple case of four oscillators. We then provided an analytical expression (involving the relative angle of squeezing, squeezing parameter, and coupling constant) that captures all aspects of the entanglement localization. Our results may be useful in situations where one wishes to control entanglement generation and distribution in solid state systems where the distance between the coupled quantum systems is so small that the use of flying qubits (typically photons) is not appropriate. In these cases, besides local control of the initial state of each quantum system, one has in general to appeal to the external control of the coupling constants to process quantum-information. Here, we provide a way to generate and localize entanglement that does not involve external control of the coupling constants but only local control of the individual (not entangled) initial quantum states. Finally, we suggested a system consisting of capacitively coupled microelectromechanical resonators to implement the ideas presented in this paper.
\\

We would like to thank J. P. Santos for helping with the figures. This work was partially supported by CNPq (Conselho Nacional de Desenvolvimento Cient\'\i fico e Tecnol\'ogico) (F.L.S.).

\end{document}